\documentclass{elsart}

\usepackage{graphicx}
\usepackage{amssymb}

\begin{document}

\begin{frontmatter}

\title{Synchronization of phase oscillators with heterogeneous coupling: A solvable case}

\author{Gabriel H. Paissan}
\ead{paissan@cab.cnea.gov.ar} \and
\author{Dami\'an H. Zanette}
\ead{zanette@cab.cnea.gov.ar}

\address{Consejo Nacional de Investigaciones Cient\'{\i}ficas y
T\'ecnicas, Centro At\'omico Bariloche and Instituto Balseiro,
8400 Bariloche, R\'{\i}o Negro, Argentina}

\begin{abstract}
We consider an extension of Kuramoto's model of coupled phase
oscillators where oscillator pairs interact with different
strengths. When the coupling coefficient of each pair can be
separated into two different factors, each one associated to an
oscillator, Kuramoto's theory for the transition to synchronization
can be explicitly generalized, and the effects of coupling
heterogeneity on synchronized states can be analytically studied.
The two factors are respectively interpreted as the weight of the
contribution of each oscillator to the mean field, and the coupling
of each oscillator to that field. We explicitly analyze the effects
of correlations between those weights and couplings, and show that
synchronization can be completely inhibited when they are strongly
anti-correlated. Numerical results validate the theory, but suggest
that finite-size effect are relevant to the collective dynamics
close to the synchronization transition, where oscillators become
entrained in synchronized frequency clusters.
\end{abstract}

\begin{keyword}
coupled oscillators \sep synchronization \sep frequency clustering
\PACS 05.45.Xt \sep 89.75.Fb \sep 05.70.Fh
\end{keyword}
\end{frontmatter}

\section{Introduction}

Macroscopic periodic oscillations are ubiquitous in living
organisms, and span a broad range of time scales, from fractions of
a second --such as in neural and heart tissues-- to days or weeks
--such as in circadian rhythms and menstrual cycles. The possibility
that this kind of macroscopic dynamics is the collective
manifestation of mutual synchronization of microscopic oscillations
was advanced by N. Wiener in the 1940s \cite{Wiener}, and was later
elaborated by A. T. Winfree \cite{Winfree68,Winfree}. The basic
mechanism assumed to be at work in this self-organization phenomenon
is that a few elements in a population of interacting oscillators
may synchronize if they have similar frequencies and their coupling
is strong enough. Under suitable conditions, other oscillators may
in turn be entrained by this ``nucleation center'' and, eventually,
contribute to form a macroscopic oscillating cluster. Collective
oscillations would thus emerge as the result of coherent microscopic
dynamics, through a process not unlike a condensation phase
transition  \cite{Winfree68}.

This conceptual idea was realized into an explicit model by Y.
Kuramoto \cite{kura}, who considered the collective dynamics of an
ensemble of $N$ globally coupled phase oscillators. The state of
each oscillator is defined by its phase  $\phi_i (t) \in [0,2\pi)$.
Their joint evolution is governed by the equations
\begin{equation} \label{kura1}
\dot \phi_i = \omega_i + \frac{K}{N} \sum_{j=1}^N  \sin
(\phi_j-\phi_i) ,
\end{equation}
($i=1,\dots,N$) where $\omega_i$ is the natural frequency of
oscillator $i$, and $K$ is the strength of their global (all-to-all)
coupling. In this system, synchronized states can be characterized
by the distribution of effective frequencies $\omega_i'$. The
effective frequency of an oscillator is defined as the time average
of its phase velocity:
\begin{equation} \label{omp}
\omega_i' = \lim_{T\to \infty}\frac{1}{T} \int_0^T \dot \phi_i dt.
\end{equation}
It is found that, for sufficiently large $K$, there is a group of
oscillators whose effective frequencies are identical. The number of
these synchronized oscillators increases with $K$. Synchronization
implies as well a certain degree of order in the distribution of the
individual oscillator phases. This is revealed by the complex order
parameter
\begin{equation} \label{zk}
z(t)\equiv \sigma (t) \exp [ {\rm i} \Phi (t)] = \frac{1}{N}
\sum_{j=1}^N \exp \left[ {\rm i} \phi_j(t) \right]  .
\end{equation}
Kuramoto showed that in the limit $N \to \infty$, and under the
assumption that synchronized oscillators form a single group with
effective frequency $\omega_i' = \Omega$, the collective amplitude
$\sigma$ --which turns out to be independent of time-- vanishes for
coupling intensities below a certain critical value $K_c$, and is
different from zero for $K>K_c$ \cite{kura}. The threshold $K_c$ of
the synchronization transition is determined by the distribution of
natural frequencies $\omega_i$: ensembles with wider frequency
distributions require stronger coupling to become synchronized. As a
byproduct of the formulation, the synchronization frequency $\Omega$
is also obtained, and  $\Phi (t) = \Phi(0)+\Omega t$.

The quantity $z(t)$ can be interpreted as a mean field generated by
the oscillator ensemble. In fact, $\sigma (t)$ and $\Phi(t)$ define,
respectively, the collective amplitude and phase of the ensemble.
The evolution equation (\ref{kura1}) for each oscillator can be
rewritten as
\begin{equation}
\dot \phi_i =\omega_i + K \sigma \sin (\Phi-\phi_i).
\end{equation}
The coupling of oscillator $i$ with the ensemble is thus represented
as the interaction with a single, equivalent oscillator of phase
$\Phi$, weighted by the amplitude $\sigma$.

The synchronization transition results from the competing effect of
two factors: while coupling induces the emergence of ordered
dynamics, the heterogeneity in the distribution of natural
frequencies favours incoherent behaviour \cite{mmz}. The aim of the
present paper is to incorporate two additional sources of
heterogeneity to Kuramoto's model. First, we consider that each
oscillator $i$ is coupled to the mean field $z=\sigma \exp ({\rm i}
\Phi)$ with its own coupling intensity $k_i$. Second, we assume that
the individual contribution of each oscillator to the mean field is
weighted by a factor $q_i$. In contrast with natural frequencies,
which determine the individual dynamics in the absence of coupling,
the two new attributes $k_i$ and $q_i$ affect the way in which each
oscillator interacts with the ensemble. Therefore, they define a
heterogeneous interaction pattern underlying the system.

Heterogeneity is a widespread feature in multi-agent systems.
Constitutional differences between the members of such systems imply
variations in the individual dynamics, in the response to external
influxes, in the interaction with other elements, etc. Those
differences may also have non-trivial effects on the collective
dynamics of the ensemble \cite{Winfree,mcal}. In connection with a
more realistic representation of both natural and artificial
systems, it is therefore desirable to generalize standard models in
order to encompass heterogeneity. The present extension of the
phase-oscillator model has the additional interest of being
analytically tractable through a rather straightforward
generalization of Kuramoto's theory. In fact, in the limit of an
infinitely large ensemble, the synchronization transition, as well
as the collective dynamical properties of the synchronized state,
can be fully characterized in the frame of such generalization.

In the next section, we discuss how Kuramoto's theory is extended to
include the form of heterogeneous coupling advanced above. We
establish self-consistency equations for the collective amplitude
and the synchronization frequency, and find the frequency
distribution of non-synchronized oscillators. In Section
\ref{sect2}, the results of Section \ref{sect1} are applied to
analyze synchronization properties in several specific cases, paying
special attention to the effect of correlations between the
different attributes of individual oscillators. In particular, we
show that anti-correlation between $k_i$ and $q_i$ can completely
suppress synchronized states. Numerical results for finite systems
are presented in Section \ref{num}, which validate analytical
predictions and, at the same time, show that finite-size effects
play an important role in the collective dynamics close to the
synchronization transition. In this region, partial synchronization
in the form of frequency clustering --not predicted by the
analytical approach-- is observed. Finally, our main results are
summarized and commented in the last section.

\section{Synchronization transition for heterogeneous coupling}

\label{sect1}

Consider the collective dynamics of a population of $N$ phase
oscillators, whose phases $\phi_i (t)$ evolve according to
\begin{equation} \label{eq1}
\dot \phi_i = \omega_i + \frac{1}{N} \sum_{j=1}^N W_{ij} \sin
(\phi_j-\phi_i) ,
\end{equation}
$i=1,\dots , N$. The coefficients $W_{ij}\ge 0$ weight the
interaction of each oscillator pair. We assume that these
coefficients can be expressed in the form of a product, $W_{ij} =k_i
q_j$, with $k_i>0$ and $q_i \ge 0$ for all $i$. Equations
(\ref{eq1}) can be written as
\begin{equation} \label{macro}
\dot \phi_i = \omega_i + k_i \sigma \sin (\Phi-\phi_i)
\end{equation}
where $\sigma(t)$ and $\Phi(t)$ are the modulus and phase of the
complex number [cf. Eq. (\ref{zk})]
\begin{equation}  \label{mfield}
z(t) \equiv \sigma (t) \exp [{\rm i} \Phi(t)] = \frac{1}{N}
\sum_{j=1}^n q_j \exp [{\rm i}\phi_j(t)] .
\end{equation}
Kuramoto's model, described in the Introduction, is recovered for
$k_i=K$ and $q_i=1$ for all $i$. The special case $k_i=q_i\equiv
s_i$ for all $i$ was studied by H. Daido \cite{Daido1,Daido2}.
However, inspired by disordered spin systems, he admitted $s_i$ to
be positive or negative.

Equation (\ref{macro}) shows that, as in Kuramoto's model, the
evolution of each oscillator can be though of as resulting from its
interaction with the mean field $z$, generated  by the oscillator
ensemble. The factor $k_i$ can be interpreted as the coupling of
oscillator $i$ with the mean field. Meanwhile, $q_i$ weights the
contribution of the same oscillator to $z$. In view of this, we
require that $q_i$ satisfies the normalization condition
\begin{equation} \label{norm}
\frac{1}{N} \sum_{i=1}^n q_i =1.
\end{equation}
This choice does not imply loosing generality: if $N^{-1} \sum_i q_i
= Q \neq 1$, we just redefine $q_i/Q \to q_i$ and $Qk_i \to k_i$,
leaving the evolution equations invariant.

We focus on the limit of an infinitely large ensemble, $N \to
\infty$, and describe the system in terms of continuous
distributions. Specifically, the distribution of individual natural
frequencies $\omega_i$, couplings $k_i$, and weights $q_i$, is given
by a function $G(\omega,k,q)$. This distribution is normalized to
unity,
\begin{equation}
\int_{-\infty}^\infty d\omega  \int_0^\infty dk  \int_0^\infty dq  \
G(\omega,k,q) =1,
\end{equation}
and, according to Eq. (\ref{norm}), it must satisfy the additional
condition
\begin{equation} \label{addit}
\int_{-\infty}^\infty d\omega  \int_0^\infty dk  \int_0^\infty dq  \
q \ G(\omega,k,q) =1.
\end{equation}
As for the distribution of individual phases, the quantity $n (\phi,
t; k,q) d\phi \ dk\ dq$ represents the fraction of oscillators with
coupling in $(k,k+dk)$ and weight in $(q,q+dq)$, whose phases lie in
$(\phi,\phi+d\phi)$ at time $t$. In terms of the density $n (\phi,
t; k,q)$, Eq. (\ref{mfield}) reads
\begin{equation} \label{sigc}
\sigma \exp({\rm i} \Phi) = \int_0^\infty dk   \int_0^\infty q\ dq
\int_0^{2\pi} d\phi \ n (\phi,t;k,q) \exp (i\phi) .
\end{equation}

As in Kuramoto's theory, we assume that the density $n (\phi,t;k,q)$
has two contributions, $n=n_u+n_s$. The first contribution
represents non-synchronized oscillators, which are uniformly
distributed in phase, so that their density $n_u$ does not depend on
$\phi$. From Eq. (\ref{sigc}), it is clear that non-synchronized
oscillators do not contribute to the mean field, because the
integration over $\phi$ vanishes.

The second contribution represents a group of synchronized
oscillators, which are assumed to possess a constant collective
frequency $\Omega$. Their density has the form $n_s (\phi,t;k,q)
\equiv n_s (\phi-\Omega t;k,q)$. Replacing this in Eq. (\ref{sigc}),
the macroscopic phase turns out to be $\Phi (t)= \Phi(0) +\Omega t$,
and
\begin{equation} \label{mfields}
\sigma \exp[{\rm i} \Phi(0)] = \int_0^\infty dk   \int_0^\infty q\
dq \int_0^{2\pi} d\psi \ n_s (\psi;k,q) \exp (i\psi) .
\end{equation}
Within this Ansatz, thus, $\sigma$ is independent of time. With an
appropriate choice of the origin of phases, we can fix $\Phi(0)=0$.

Introducing the macroscopic phase $\Phi (t)= \Omega t$ in Eq.
(\ref{macro}), and defining the individual deviation from $\Phi$ as
$\psi_i(t)=\phi_i (t) -\Phi (t)$, we get
\begin{equation} \label{psi}
\dot \psi_i = \omega_i-\Omega - k_i \sigma \sin  \psi_i.
\end{equation}
In this equation for $\psi_i$, the collective amplitude $\sigma$ and
the synchronization frequency $\Omega$ are not known. However, we do
know that, in the present formulation, they are independent of time.
Therefore, Eq. (\ref{psi}) can be formally solved for generic
constant values of $\sigma$ and $\Omega$. Once the deviations
$\psi_i$ --and, thus, the distribution of phases-- have been found,
the unknowns $\sigma$ and $\Omega$ are calculated from Eq.
(\ref{sigc}). This self-consistent calculation of the collective
properties of the oscillator ensemble is the core of Kuramoto's
theory.

When $|\omega_i -\Omega|\le k_i \sigma $, Eq. (\ref{psi}) has a
stable fixed point at
\begin{equation} \label{relac}
\psi_i = \arcsin \frac{\omega_i -\Omega}{k_i \sigma},
\end{equation}
with $-\pi/2 \le \psi_i \le \pi/2$. This fixed point stands for the
(time-independent) asymptotic phase deviation of a synchronized
oscillator of natural frequency $\omega_i$ and coupling $k_i$ with
respect to the macroscopic phase $\Phi (t)=\Omega t$.
Asymptotically, the effective frequency of synchronized oscillators,
Eq. (\ref{omp}), is $\omega_i'=\Omega$. For each value of the
coupling, Eq. (\ref{relac}) relates the asymptotic phase of a
synchronized oscillator with its natural frequency. Consequently,
this equation can be used to link the asymptotic density $n_s (\psi;
k,q)$ of synchronized oscillators to the distribution $G(\omega,k,q
)$ of natural frequencies, couplings and weights. Taking into
account the relation
\begin{equation}
n_s (\psi; k,q)\ d\psi \ dk \ dq = G(\omega,k,q )\ d\omega\  dk\ dq,
\end{equation}
which holds for $-\pi/2 \le \psi \le \pi/2$, and noting that Eq.
(\ref{relac}) implies the differential relation $ k \sigma \cos \psi
\ d \psi \ dk \ dq = d \omega \ dk \ dq$, we find
\begin{equation} \label{ns}
n_s (\psi;k,q) =\left\{
\begin{array}{ll}
k\sigma G( \Omega+k\sigma\sin \psi,k,q) \cos \psi  & \mbox{   for
$-\pi/2 < \psi < \pi/2$,} \\ \\ 0 & \mbox{   otherwise.} \end{array}
\right.
\end{equation}
Replacing this form of the density of synchronized oscillators in
Eq. (\ref{mfields}), and separating real and imaginary parts, yields
the self-consistency equations
\begin{equation} \label{12}
\sigma = \sigma \int_0^\infty k\ dk   \int_0^\infty q\ dq
\int_{-\pi/2}^{\pi/2} d\psi \ G( \Omega+k\sigma\sin \psi,k,q) \cos^2
\psi ,
\end{equation}
and
\begin{equation} \label{13}
0 = \sigma \int_0^\infty k\ dk   \int_0^\infty q\ dq
\int_{-\pi/2}^{\pi/2} d\psi \ G( \Omega+k\sigma\sin \psi,k,q) \cos
\psi \sin \psi ,
\end{equation}
for the  collective amplitude $\sigma$ and the synchronization
frequency $\Omega$, in terms of the distribution $G(\omega,k,q)$.
Once these quantities are obtained, the density of synchronized
oscillators can be explicitly calculated using Eq. (\ref{ns}). We
postpone the discussion of the solutions to Eqs. (\ref{12}) and
(\ref{13}), which define the synchronization properties of the
oscillator ensemble, to the next section.

Oscillators for which $|\omega_i -\Omega|> k_i \sigma$ do not reach
a stationary deviation from the macroscopic phase $\Phi$ and,
therefore, do not synchronize. In this case, the solution to Eq.
(\ref{psi}) implies that the phase of a non-synchronized oscillator
varies with time as
\begin{equation}
\phi_i (t) = \omega_i' t + \xi_i [(\omega_i-\Omega)t],
\end{equation}
where $\xi_i (t)$ is a $2\pi$-periodic function of $t$. The
effective frequency $\omega_i'$ is given by
\begin{equation} \label{omef}
\omega_i'=\Omega+(\omega_i-\Omega) \sqrt{1- \left(
\frac{k_i\sigma}{\omega_i-\Omega} \right)^2}.
\end{equation}
For each value of $k_i$, this equation links the effective frequency
$\omega_i'$ with the natural frequency $\omega_i$ of each
non-synchronized oscillator. It makes it possible to calculate the
distribution $G'(\omega',k,q)$ of effective frequencies, couplings
and weights, in terms of $G(\omega,k,q)$. Taking into account the
relation $G'(\omega',k,q) d\omega' \ dk \ dq = G(\omega,k,q) d\omega
\ dk \ dq $, and using Eq. (\ref{omef}) to obtain the ratio $d\omega
/d\omega '$, we get
\begin{equation}
G'(\omega',k,q)=
\frac{|\omega'-\Omega|}{\sqrt{(\omega'-\Omega)^2+k^2\sigma^2}} \
 G\left[ \Omega+\sqrt{(\omega'-\Omega)^2+k^2\sigma^2},k,q \right] .
\end{equation}
This distribution of effective frequencies for non-synchronized
oscillators can be explicitly calculated once $\sigma$ and $\Omega$
have been found.

\section{Analysis of synchronization properties}

\label{sect2}

The self-consistency equations (\ref{12}) and (\ref{13}) define the
collective macroscopic properties of the synchronized oscillators in
terms of the distribution $G(\omega,k,q)$ of individual frequencies,
couplings, and weights. A full analysis of the solutions for an
arbitrary form of $G(\omega,k,q)$, taking into account any possible
dependence on its three variables, is out of reach. In the following
subsections, consequently, we study in detail a few representative
special cases.

It is however possible to begin by pointing out a few generic
properties of Eqs. (\ref{12}) and (\ref{13}). First of all, these
equations --and, as a matter of fact, the full formulation presented
in Sect. \ref{sect1}-- reduces to Kuramoto's theory when all
oscillators have the same coupling $K$ and unitary weight, i.e. when
\begin{equation}
G(\omega,k,q) \equiv g(\omega) \delta (k - K) \delta (q - 1),
\end{equation}
as expected. In this limit, the only relevant individual attribute
is the natural frequency $\omega_i$, with distribution $g(\omega)$.

For any form of $G(\omega,k,q)$, a trivial solution to Eqs.
(\ref{12}) and (\ref{13}) is $\sigma=0$. According to Eq.
(\ref{ns}), this corresponds to a fully non-synchronized collective
state. To study synchronization features we look for solutions with
non-vanishing mean field, $\sigma\neq 0$. Moreover, as is known to
happen in Kuramoto's theory, if the distribution $G(\omega,k,q)$ is
an even function with respect to the variable $\omega$ around a
certain value $\omega_0$, i.e.
\begin{equation}
G(\omega_0+\omega,k,q)=G(\omega_0-\omega,k,q),
\end{equation}
Eq. (\ref{13}) is satisfied by taking $\Omega=\omega_0$. In other
words, for a symmetric distribution of natural frequencies, the
synchronization frequency coincides with the center of the
distribution. Now, since in Eqs. (\ref{eq1}) natural frequencies are
defined up to an arbitrary additive constant (corresponding to a
phase rotation at constant angular velocity) we can always choose
$\omega_0=0$. In this case, the synchronization frequency vanishes.
Under the above conditions, the problem reduces to Eq. (\ref{12}) in
the form
\begin{equation} \label{12a}
1 = \int_0^\infty k\ dk   \int_0^\infty q\ dq \int_{-\pi/2}^{\pi/2}
d\psi \ G( k\sigma\sin \psi,k,q) \cos^2 \psi.
\end{equation}
The synchronization threshold, at which the non-trivial solution
reaches its lowest value $\sigma=0$, is defined by the condition
\begin{equation} \label{12aa}
\int_0^\infty k\ dk \int_0^\infty q \ G(0,k,q) \ dq = \frac{2}{\pi}
 .
\end{equation}
In the following, for the sake of conciseness, we deal with this
specific situation.

Before passing to the consideration of Eqs. (\ref{12a}) and
(\ref{12aa}) for some special forms of $G(\omega,k,q)$, let us point
out an important property regarding the dependence of our problem on
the variable $q$. Suppose that the weights $q_i$ are not correlated
with the natural frequencies $ \omega_i$ or with the couplings
$k_i$, so that we have
\begin{equation}
G(\omega,k,q) \equiv G_1 (\omega,k) G_2 (q).
\end{equation}
Then, the distribution of weights becomes irrelevant to the problem.
Indeed, under such condition, the integration over $q$ in Eqs.
(\ref{12a}) and (\ref{12aa}) --as well as in Eqs. (\ref{12}) and
(\ref{13})-- can be trivially performed, taking into account that
Eq. (\ref{norm}) requires
\begin{equation}
\int_0^\infty q \ G_2(q) \ dq = 1.
\end{equation}
From then on, the distribution of weights plays no role in defining
the macroscopic synchronization properties of the ensemble. In the
density of synchronized oscillators, Eq. (\ref{ns}), $G_2(q)$
appears as a trivial factor modulating the shape of $n_s
(\psi;k,q)$. This weak role of the variable $q$ in the solution to
the present problem is explained by noting that, in an infinitely
large ensemble, an average quantity such as the mean field $z$ does
not change if each term in the average is weighted by a random
coefficient, as long as the mean value of these coefficients equals
unity. In our formulation, this property can be traced back to Eq.
(\ref{macro}), which governs the evolution of each oscillator. Once
the mean field $z$ has been introduced, the individual weights $q_i$
disappear from the calculation.

This situation changes drastically if, on the other hand, there is a
correlation between the  weight and the coupling or the natural
frequency of each oscillator. In this case, since the asymptotic
phase of each synchronized oscillator depends on its coupling and
frequency as given by Eq. (\ref{relac}), a correlation emerges
between the weight and the phase. As a consequence, the mean field
depends on how $q$ is distributed, and on its correlation with
$\omega$ and $k$. In view of this remark, our analysis of
non-trivial distributions of $q$ will focus on the case where the
attributes $\omega_i$, $k_i$, and $q_i$ of each individual
oscillator are mutually correlated.

\subsection{Uniform weights}

We address first the case where the weights are equal for all
oscillators, $q_i =1 $ for all $i$. In this situation,
$G(\omega,k,q) \equiv G_1 (\omega,k) \delta (q-1) $, and Eq.
(\ref{12a}) becomes
\begin{equation} \label{12b}
1 = \int_0^\infty k\ dk   \int_{-\pi/2}^{\pi/2} d\psi \ G_1
(k\sigma\sin \psi,k) \cos^2 \psi.
\end{equation}
In Ref. \cite{nos}, we performed a preliminary analysis of Eq.
(\ref{12b}), paying particular attention to the case where couplings
and natural frequencies are in turn uncorrelated, $G_1
(\omega,k)\equiv g(\omega) h(k)$. There, it was found that the
synchronization threshold is given by the condition
\begin{equation} \label{kcrit1}
\langle k \rangle \equiv \int_0^\infty  k \ h(k) \ dk = \frac{2}{\pi
g(0)}.
\end{equation}
In Kuramoto's theory, where all couplings are equal, $k_i = K =
\langle k \rangle$ for all $i$, the critical value of the coupling
at which synchronization switches on is $K_c = 2/\pi g(0)$.
Therefore, Eq. (\ref{kcrit1}) establishes that, in the case where
couplings are distributed (and uncorrelated to natural frequencies),
synchronization becomes possible when the average coupling  reaches
Kuramoto's threshold $K_c$. Close to the synchronization transition,
the collective amplitude behaves as
\begin{equation}
\sigma = \frac{4}{\sqrt{\pi K_c \langle k^3 \rangle |g''(0)|}}
(\langle k \rangle -K_c)^{1/2} ,
\end{equation}
where $g''(0)$ is the second derivative of the distribution of
natural frequencies at the center of the distribution, $\omega=0$.
As in Kuramoto's theory, this is assumed to be a negative number,
corresponding to a maximum in the distribution.

More generally, if the joint distribution of natural frequencies and
couplings behaves as $G_1 (\omega,k) = G_0 (k) -\gamma (k) \omega^2$
close to $\omega=0$,  the synchronization transition takes place
when
\begin{equation} \label{kcrit2}
 \int_0^\infty  k \ G_0 (k) \ dk = \frac{2}{\pi }.
\end{equation}
Close to the transition, the solution of Eq. (\ref{12b}) is
approximately given by
\begin{equation}
\sigma =\sqrt{ \frac{ \frac{\pi}{2} \int_0^\infty k G_0(k)  dk
-1}{\frac{\pi}{8} \int_0^\infty k^3 \gamma (k)  dk} } .
\end{equation}
The possibility of having synchronization is determined both by the
number of oscillators with natural frequencies at the center of the
distribution and by their coupling distribution. In fact, Eq.
(\ref{kcrit2}) can be rewritten as
\begin{equation}
\langle k \rangle_0 \ g_0 = \frac{2}{\pi },
\end{equation}
where $\langle k \rangle_0$ is the average coupling of the
oscillators with $\omega=0$, and $g_0 = \int_0^\infty G_0 (k) dk$ is
the density of these oscillators. To become synchronized, a low
number of oscillators at the center of the distribution requires
that their average coupling is high, and {\it vice versa}.

\subsection{Uniform couplings}

When the couplings of all oscillators are equal, $k_i=K$ for all
$i$, the distribution of natural frequencies, couplings, and weights
is factorized as $G(\omega,k,q) \equiv G_2 (\omega,q) \delta (k-K)$,
and Eq. (\ref{12a}) reads
\begin{equation} \label{12c}
1 = K  \int_0^\infty q\ dq \int_{-\pi/2}^{\pi/2} d\psi \ G_2( K
\sigma\sin \psi , q) \cos^2 \psi.
\end{equation}
As discussed above, unless the function $G_2 (\omega,q)$ establishes
a correlation between weights and natural frequencies, the problem
would be equivalent to Kuramoto's case.

Even if there is a correlation between weights and natural
frequencies, this case of uniform couplings has the same
mathematical form as Kuramoto's problem. In fact, Eq. (\ref{12c})
can be rewritten as
\begin{equation} \label{12c1}
1 = K  \int_{-\pi/2}^{\pi/2} d\psi \ \tilde g ( K \sigma\sin \psi)
\cos^2 \psi,
\end{equation}
with
\begin{equation} \label{gtilde}
\tilde g (\omega) = \int_0^\infty q\ G_2(\omega , q) \ dq .
\end{equation}
Equation (\ref{12c1}) coincides with the equation for $\sigma$ of an
ensemble with uniform couplings and weights, and with an effective
frequency distribution $\tilde g (\omega)$.

To evaluate the effect of correlations between weights and natural
frequencies, we consider the extreme case where the weight $q_i$ of
each oscillator is given by a function of its frequency $\omega_i$,
say, $q_i = \theta (\omega_i)$. In this case, $G_2 (\omega, q) =
g(\omega) \delta [q-\theta(\omega)]$.  Equation (\ref{addit})
requires $\int_{-\infty}^\infty g(\omega)\theta(\omega) d \omega =1
$. The effective distribution of frequencies in Eq. (\ref{12c1})
turns out to be $ \tilde g (\omega) = \theta (\omega) \  g
(\omega)$, so that the condition for the synchronization transition
reads
\begin{equation}
K = \frac{2}{\pi \tilde g(0)} =  \frac{2}{\pi \theta(0) g(0)} .
\end{equation}
This result shows that, with respect to the case of uncorrelated
weights, the threshold coupling for synchronization decreases when
the weights of the oscillators at the center of the frequency
distribution are larger than the average, $\theta (0) >1$, i.e. when
the contribution of those oscillators to the mean field is
relatively strong. On the contrary, if the weight of those
oscillators in relatively small, a stronger coupling is necessary to
induce synchronization.

\subsection{Weight-coupling correlation} \label{wcc1}

A more interesting situation arises when both weights and couplings
adopt non-uniform values over the ensemble. In this case, as shown
below, the correlation between $k_i$ and $q_i$ can have drastic
effects on the collective behaviour of oscillators, to the point of
suppressing synchronization even in the presence of arbitrarily
strong couplings.

To simplify the discussion, we assume that individual natural
frequencies are not correlated with weights or couplings, so that
$G(\omega, k,q) = g(\omega) G_3 (k,q)$. As for the weight-coupling
correlation, we consider the extreme case in which $q_i$ is a given
function of $k_i$,  $q_i = \Theta (k_i)$, so that $G_3 (k,q) = h(k)
\delta [q-\Theta(k)] $. Condition (\ref{addit}) requires that
\begin{equation} \label{wcn}
\int_0^\infty h(k) \Theta (k) dk=1,
\end{equation}
and the threshold for synchronization is determined by the equation
\begin{equation} \label{wcn1}
\int_0^\infty k\ h(k) \ \Theta(k) \ dk = \frac{2}{\pi g(0)}.
\end{equation}

Suppose now that the product $h(k) \Theta (k)$ satisfies Eq.
(\ref{wcn}), and compare Eq. (\ref{wcn1}) with Eq. (\ref{kcrit1}),
which determines the synchronization threshold for the case of
uniform weights. If $q=\Theta (k)$ is an increasing function of $k$
--i.e., if weights and couplings are positively correlated-- and for
a given value of $g(0)$, the synchronization condition is expected
to hold in the present case for relatively lower couplings. On the
other hand, if $\Theta (k)$ decreases with $k$ --i.e., if weights
and couplings are anti-correlated-- relatively larger couplings will
be required to reach the synchronization threshold. It may even
happen, as we show below, that a sufficiently strong
anti-correlation between weights and couplings suppresses the
occurrence of synchronization even when arbitrarily large couplings
are present in the ensemble.

From the viewpoint of the dynamical roles of weights and couplings,
this effect of positive or negative correlation between $k_i$ and
$q_i$ is interpreted as follows. The emergence of synchronized
behaviour is facilitated if the oscillators whose coupling with the
mean field $z$ is stronger are in turn those whose contribution to
$z$ is relatively larger. This kind of feedback between oscillators
with large $k_i$ and $q_i$ enhances the development of coherent
states. On the other hand, if the mean field is dominated by
oscillators whose coupling with the ensemble is weak,
synchronization can be inhibited or even fully suppressed.

To illustrate these phenomena, we explicitly work out a case with a
particularly simple distribution of couplings, which is later used
again in our numerical simulations of Section \ref{num}. Let us
consider a uniform distribution, $h(k)=K_{\max}^{-1}$ for
$0<k<K_{\max}$, and $h(k)=0$ otherwise. As for the weights, we take
\begin{equation} \label{expon}
q= \Theta (k) = \frac{\lambda K_{\max}}{\exp (\lambda K_{\max})-1 }
\exp(\lambda k),
\end{equation}
where the pre-factor has been chosen to satisfy Eq. (\ref{wcn}). The
weight-coupling correlation is positive for $\lambda>0$, because $q$
grows with $k$. For $\lambda <0$, on the other hand, $q$ and $k$ are
anti-correlated. In the limit $\lambda=0$, the case of uniform
weights is reobtained.

With these choices, Eq. (\ref{wcn1}) yields
\begin{equation} \label{lamb}
 \frac{1-(1-\lambda K_{\max}) \exp (\lambda
K_{\max}) }{\lambda[\exp (\lambda K_{\max})-1]}=\frac{2}{\pi g(0)} ,
\end{equation}
which can be interpreted as an equation for the maximum coupling
$K_{\max}$ at the synchronization transition. For small $\lambda$,
the solution to this equation reads
\begin{equation}
K_{\max} \approx \frac{4}{\pi g(0)}-\frac{8 \lambda}{3\pi^2 g(0)^2}
.
\end{equation}
As expected, with respect to the case of uniform weights --where the
synchronization threshold occurs at $K_{\max} =4/\pi g(0)$-- a lower
value of $K_{\max}$ is required if weights and couplings are
positively correlated ($\lambda>0$). On the other hand, if
$\lambda<0$ and the correlation is negative, couplings must be
larger to induce synchronization.

Note now that, for negative $\lambda$, the left-hand side of Eq.
(\ref{lamb}) is an increasing function of $K_{\max}$, which
asymptotically approaches the value $|\lambda|^{-1}$ as $K_{\max}$
grows to infinity. This implies that, if $\lambda<\lambda_c=-\pi
g(0)/2$, there will be no finite value of $K_{\max}$ such that Eq.
(\ref{lamb}) is satisfied. As $\lambda$ approaches the critical
value $\lambda_c$ from above, in fact, the solution for $K_{\max}$
diverges to infinity. In other words, if the anti-correlation
between weights and couplings is strong enough, even arbitrarily
large couplings are unable to induce the synchronization transition.
For $\lambda < \lambda_c$ and any $K_{\max}$ the only possible value
for the collective amplitude is the trivial one, $\sigma=0$.

\section{Numerical results. Frequency clustering}
\label{num}

In this section, we present a series of results from the numerical
resolution of Eqs. (\ref{eq1}). The aim of these numerical
calculations is two-fold. On the one hand, they validate our main
analytical results over wide regions of the relevant parameter
spaces. On the other, they show that a systematic departure from the
analytical results is observed, in finite systems, close to the
synchronization transition. As advanced in Ref. \cite{nos}, this
departure is due to the occurrence of a regime of frequency
clustering, presumably ascribable to finite-size fluctuations, and
disregarded in the analytical approach. Part of our numerical
calculations is thus devoted to illustrate and characterize this
regime, where a portion of the ensemble segregates into several
groups of mutually synchronized oscillators. Within each group, all
oscillators share the same effective frequency.

We solve Eqs. (\ref{eq1}) by means of a standard Euler algorithm,
for ensembles of $N=10^4$ phase oscillators. In order to emphasize
differences with the case of homogeneous global coupling, we focus
on variations in the distribution of couplings $k_i$ and weights
$q_i$, and on their possible correlation, while we assume that
natural frequencies are uncorrelated to coupling and weights. The
distribution of natural frequencies is a Gaussian, $g(\omega) =
\exp(-\omega^2/2) / \sqrt{2\pi}$. Results are typically obtained
from averages over $100$ realizations of the ensemble, with
couplings, weights, and natural frequencies drawn anew from their
respective distributions.

A crucial point in our numerical calculations is the identification
of mutually synchronized oscillators or, equivalently, of
synchronized clusters. This is achieved by comparing the
distribution of natural frequencies $\omega_i$ and effective
frequencies $\omega'_i$ \cite{nos}. In a given realization of the
ensemble, natural frequencies are drawn at random from the
above-referred Gaussian distribution, and effective frequencies are
calculated from a numerical approximation to the integral in Eq.
(\ref{omp}). The distributions of $\omega_i$ and $\omega'_i$ are
constructed as histograms with columns of width $\Delta=10^{-3}$. In
the histogram of natural frequencies we identify the highest column,
whose height we denote by $h$. In our calculations, typical values
of $h$ are $10$ to $15$. A column in the histogram of effective
frequencies is considered to belong to a synchronized cluster if its
height is above $h$. A synchronized cluster is defined as a set of
contiguous columns higher than $h$ whose nearest columns to the left
and to the right are lower than $h$. Synchronized oscillators are
those oscillators belonging to synchronized clusters.

\subsection{Uncorrelated couplings and weights}

\begin{figure} \centering
\resizebox{.8\columnwidth}{!}{\includegraphics{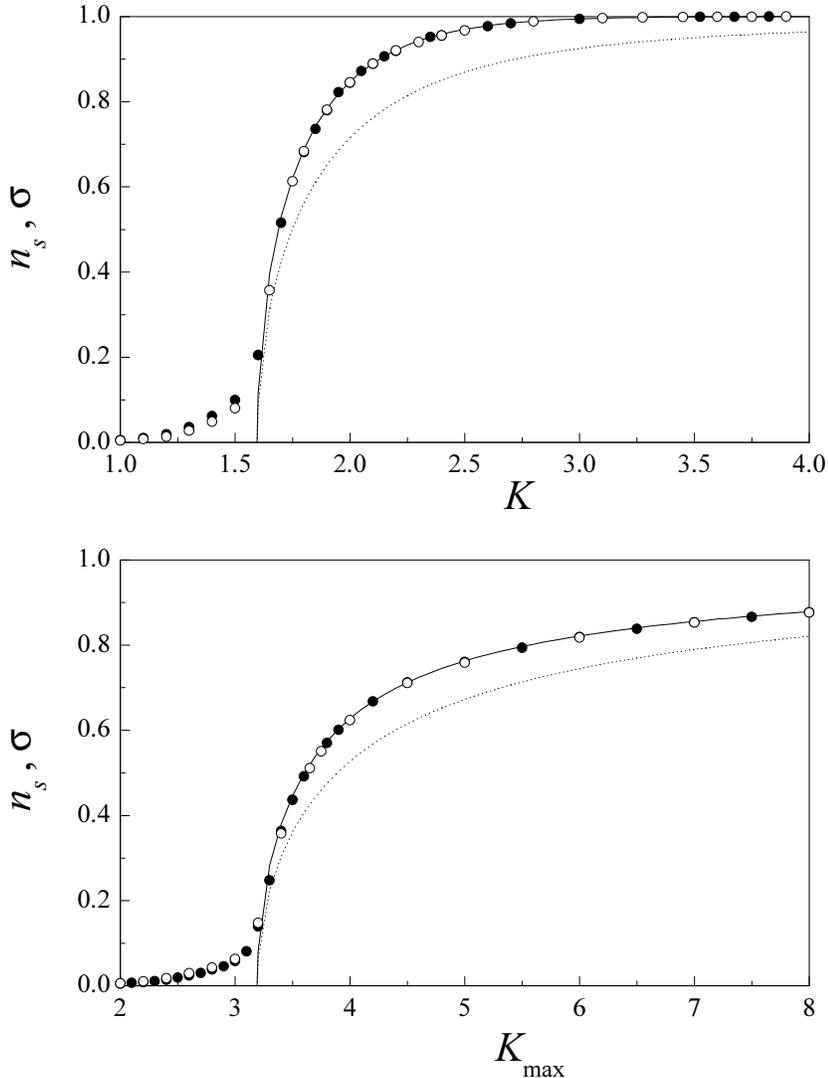}}
\caption{Upper panel: Numerical results for the fraction $n_s$ of
synchronized oscillators in the case of constant couplings, $k_i=K$
for all $i$, with constant weights, $q_i=1$ for all $i$ (solid
dots), and with uniformly distributed weights (open dots). Lower
panel: The same, for the case of uniformly distributed couplings,
$0<k_i<K_{\max}$. Full and dotted curves stand for the analytical
calculation of $n_s$ and the collective amplitude $\sigma$,
respectively.} \label{fig1}
\end{figure}

First, we have measured the total number $N_s$ of synchronized
oscillators for two uncorrelated distributions of couplings and
weights. The upper panel of Fig. \ref{fig1} shows the fraction of
synchronized oscillators, $n_s=N_s/N$, in the case of constant
couplings, $k_i=K$ for all $i$, as a function of $K$. Solid dots
correspond to the case of constant weights, $q_i=1$ for all $i$.
This case coincides with Kuramoto's model, which we use as a
reference system. Open dots, on the other hand, stand for ensembles
where the weights are drawn at random from a uniform distribution
over the interval $(0,1)$, and then normalized to satisfy Eq.
(\ref{norm}).

The full curve represents the analytical estimate for the fraction
of synchronized oscillators, defined from our analytical results as
\begin{equation}
n_s \equiv \int_0^\infty  dk   \int_0^\infty \ dq \int_{0}^{2\pi}
d\psi \ n_s(\psi;k,q),
\end{equation}
while the dotted curve is the analytical result for the collective
amplitude $\sigma$, obtained from Eq. (\ref{12a}). Both $n_s$ and
$\sigma$ are different from zero above Kuramoto's critical coupling
$K_c = 2/\pi g(0)= 2\sqrt{2/\pi} \approx 1.596$.

The agreement between numerical and analytical results is very good
for $K>K_c$. Below the synchronization transition, on the other
hand, we find a noticeable difference which, as advance above, is
associated with the formation of synchronized clusters. This
phenomenon is illustrated later. Note also that, in agreement with
our discussion on the irrelevance of the distribution of weights
when they are uncorrelated to couplings, solid and open dots above
the transition fall over the same curve.

The lower panel of Fig. \ref{fig1} displays analogous results for
the case where couplings are uniformly distributed in the interval
$(0,K_{\max})$. The corresponding distribution is $h(k)=
K_{\max}^{-1}$ for $0<k<K_{\max}$, and $h(k)=0$ otherwise, and the
synchronization transition takes place at $K_{\max}=2K_c \approx
3.192$.  Again, the agreement between numerical and analytical
results is very good above the transition, and the same kind of
deviation is observed below it. Also, as expected, the distribution
of weights $q_i$ has no effect on the fraction of synchronized
oscillators.

\begin{figure} \centering
\resizebox{.8\columnwidth}{!}{\includegraphics{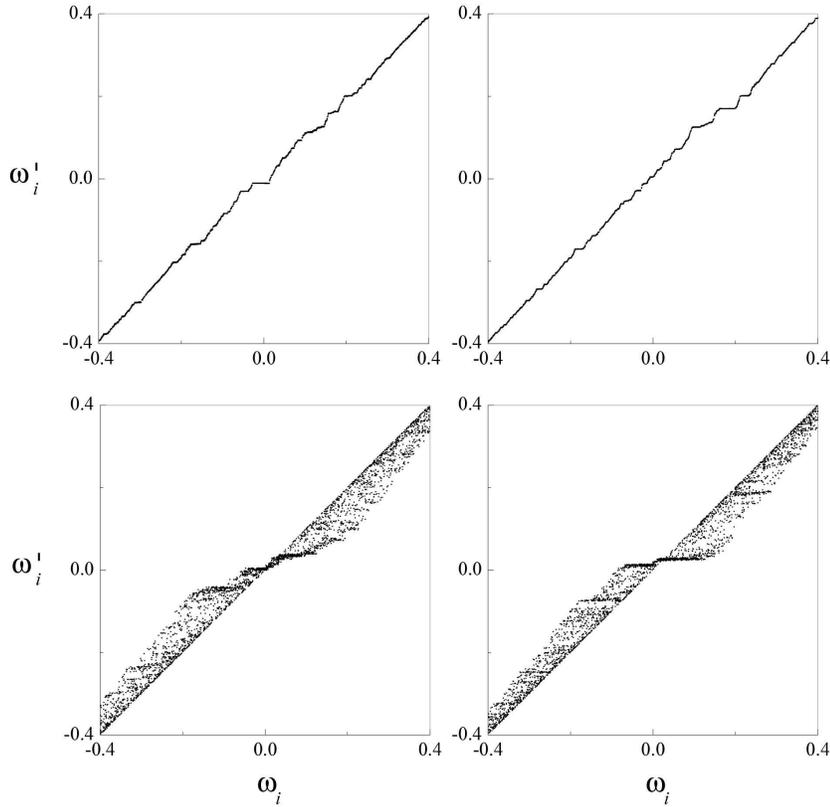}} \caption{The
effective frequency $\omega_i'$ as a function of the natural
frequency $\omega_i$ of individual oscillators in single
realizations for constant coupling, $k_i=1.5$ for all $i$, with
constant weights (upper right), and uniformly distributed weights
(upper left), and for uniformly distributed couplings, $0<k_i<3$,
with constant weights (lower right), and uniformly distributed
weights (lower left). Only the central region of the frequency
distributions is displayed.} \label{fig2}
\end{figure}

A suitable illustration of the state of the system close to the
synchronization transition, where analytical and numerical results
differ from each other, is provided by a plot of the effective
frequencies $\omega_i'$ versus the natural frequencies $\omega_i$ of
all oscillators in a given realization of the ensemble. In this kind
of plot, a horizontal array of dots --corresponding to oscillators
with different natural frequencies but the same effective
frequency-- reveals the presence of a synchronized cluster. Figure
\ref{fig2} displays such plots for the four cases already considered
in Fig. \ref{fig1}. In the upper line, couplings are constant,
$k_i=K=1.5$ for all $i$, while in the lower line they are uniformly
distributed between $0$ and $K_{\max} =3$. In both cases, thus, the
system is just below the synchronization threshold. The appearance
of synchronized clusters near the center of the frequency
distribution is apparent in the four plots.

Figure \ref{fig2} shows again that, for a given distribution of
couplings, there is little qualitative difference between the cases
where the weights $q_i$ are either constant or distributed. On the
other hand, we find that for oscillators with a given natural
frequency, the effective frequencies are essentially identical in
the case of constant couplings, but exhibit a substantial dispersion
for distributed couplings. This difference can be understood by
extrapolating our analytical results for the synchronization regime
to the present situation, just below the transition. In fact, Eq.
(\ref{omef}) implies that, for a given natural frequency, the
effective frequency depends on $k_i$, so that we expect to have the
same value of $\omega_i'$ in the case of uniform coupling, and
different values if couplings are distributed. It also predicts that
the effective frequency of a given oscillator should be smaller than
its natural frequency when the latter is larger than the
synchronization frequency $\Omega$, and {\it vice versa}. We see
from Fig. \ref{fig2} that our numerical calculations, for which
$\Omega\approx 0$, agree with such prediction.

Our previous results for the case of constant weights \cite{nos},
which we do not reproduce here, suggest that the regime of frequency
clustering is a finite-size effect due to fluctuations in the
distribution of natural frequencies. An unusual accumulation of
oscillators in a given frequency interval can trigger local
entrainment for couplings below the transition to synchronization.
Numerical results show that clustering becomes less important, with
a smaller fraction of entrained oscillators, as the ensemble grows
in size. In the limit of an infinitely large system, fluctuations
should average out and synchronization should switch on at the
analytically predicted threshold, with the formation of a
macroscopic cluster. However, the fact the frequency clustering is
still conspicuous in a relatively large ensemble as in our
calculations seems to point out that this regime plays a relevant
role in the collective dynamics of finite systems.

\subsection{Weight-coupling anti-correlation}

\begin{figure} \centering
\resizebox{.8\columnwidth}{!}{\includegraphics{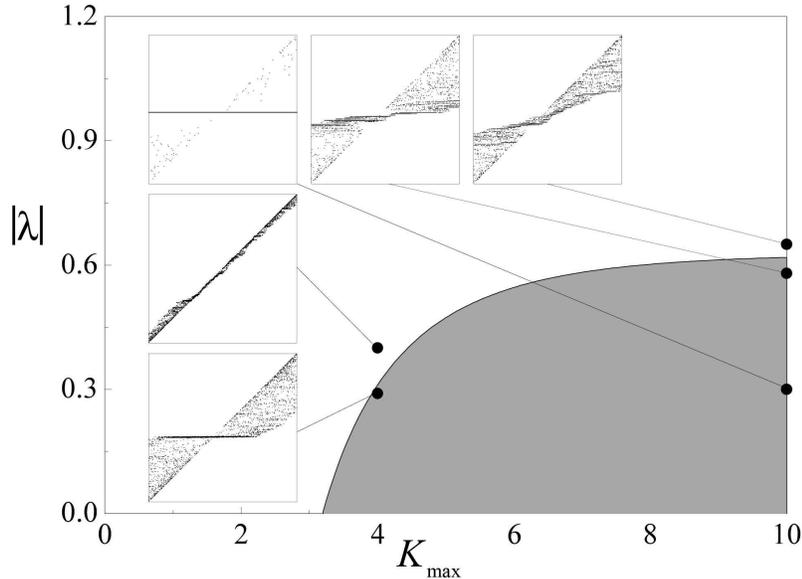}}
\caption{Synchronization region (gray shaded), obtained analytically
for anti-correlated couplings and weights, Eq. (\ref{expon}), in the
$(K_{\max},|\lambda|)$ parameter plane. The inserts show plots of
individual effective frequencies versus natural frequencies (cf.
Fig. \ref{fig2}), for single realizations of the system at the
indicated points of the parameter plane. Scales in all the inserts
vary from $-0.4$ to $0.4$ on both axes.} \label{fig3}
\end{figure}

We turn now the attention to the case discussed in Section
\ref{wcc1}, where couplings and weights are correlated. As above,
couplings are uniformly distributed on $(0,K_{\max})$, and the
weight of each oscillator is given as function of its coupling
through Eq. (\ref{expon}). We focus on the more interesting case of
weight-coupling anti-correlation, where synchronization suppression
is possible, so that we take $\lambda <0$. Results are thus
presented in terms of the positive parameter $|\lambda |$. We recall
that, in our analysis of infinitely large ensembles,  Eq.
(\ref{lamb}) predicts the relation between $\lambda$ and $K_{\max}$
at the synchronization threshold. Figure \ref{fig3} shows the region
of the $(K_{\max},|\lambda|)$ parameter plane where synchronized
states are possible. Synchronization is suppressed for small
couplings and strong weight-coupling anti-correlation. In
particular, it cannot occur for any $|\lambda|$ if $K_{\max}<2/\pi
g(0) =4\sqrt{2/\pi}\approx 3.192$, or for any $K_{\max}$ if
$|\lambda|>\pi g(0)/2= \sqrt{\pi/8} \approx 0.627$. Inserts in Fig.
\ref{fig3} illustrate the relation of effective and natural
frequencies at several points in the parameter plane.

\begin{figure} \centering
\resizebox{.8\columnwidth}{!}{\includegraphics{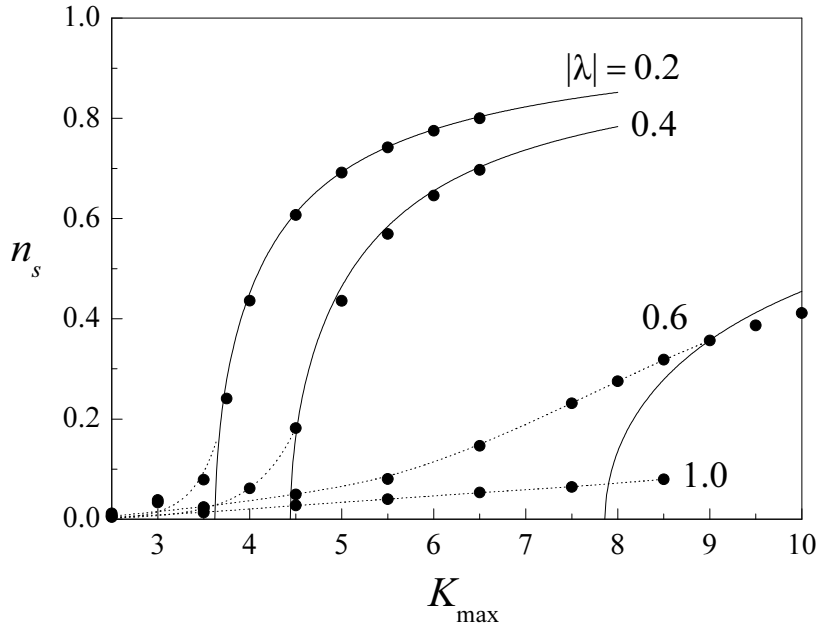}}
\caption{The fraction $n_s$ of synchronized oscillators for the case
of anti-correlated couplings and weights, as a function of
$K_{max}$, for four values of $|\lambda|$. Full curves stand for the
analytical prediction, and dotted curves are plotted as a guide to
the eye in the  region of frequency clustering.} \label{fig4}
\end{figure}

For small values of $|\lambda|$, the main effect of weight-coupling
anti-correlation is to shift the synchronization transition to
higher couplings. This is shown in Fig. \ref{fig4}, where the
fraction of synchronized oscillators is plotted as a function of
$K_{\max}$ for four values of $|\lambda|$. Full lines correspond to
the analytical prediction, which again gives a very good description
of the numerical results above the transition, at least for
$|\lambda|=0.2$ and $0.4$. Dotted lines are plotted as a guide to
the eye in the region preceding the synchronization transition. For
$|\lambda|=0.6$, the systems is practically at the limit where
synchronization becomes inhibited for any $K_{\max}$. The analytical
transition point is strongly shifted to the right, but a
considerable fraction of the ensemble is synchronized in clusters
well below the transition. Note that a discrepancy between numerical
and analytical results persists above the critical point. Finally,
for $|\lambda|=1$, synchronization should be completely suppressed
for infinitely large systems. In our numerical calculations, a small
fraction of the ensemble --which grows with $K_{\max}$-- is however
synchronized.

It turns out that the quantities that characterize synchronization
in the clustering regime --such as the fraction of synchronized
oscillators, and the number of clusters-- depend strongly on the
value of $K_{\max}$. This is illustrated, for instance, by the two
inserts of Fig. \ref{fig3} corresponding to parameters just outside
the synchronization region. For $K_{\max}=4$, clusters are small and
relatively sparse, and most effective frequencies are close to the
corresponding natural frequencies (all dots are near the diagonal).
For $K_{\max}=10$, on the other hand, a substantial fraction of the
ensemble around the center of the frequency distribution is
entrained into clusters, and form massive groups extended over
considerable intervals of natural frequencies. This effect can be
understood taking into account that a fluctuation in the
distribution of frequencies is more likely to give rise to a
synchronized cluster if the involved oscillators have, on the
average, larger couplings.

\begin{figure} \centering
\resizebox{.8\columnwidth}{!}{\includegraphics{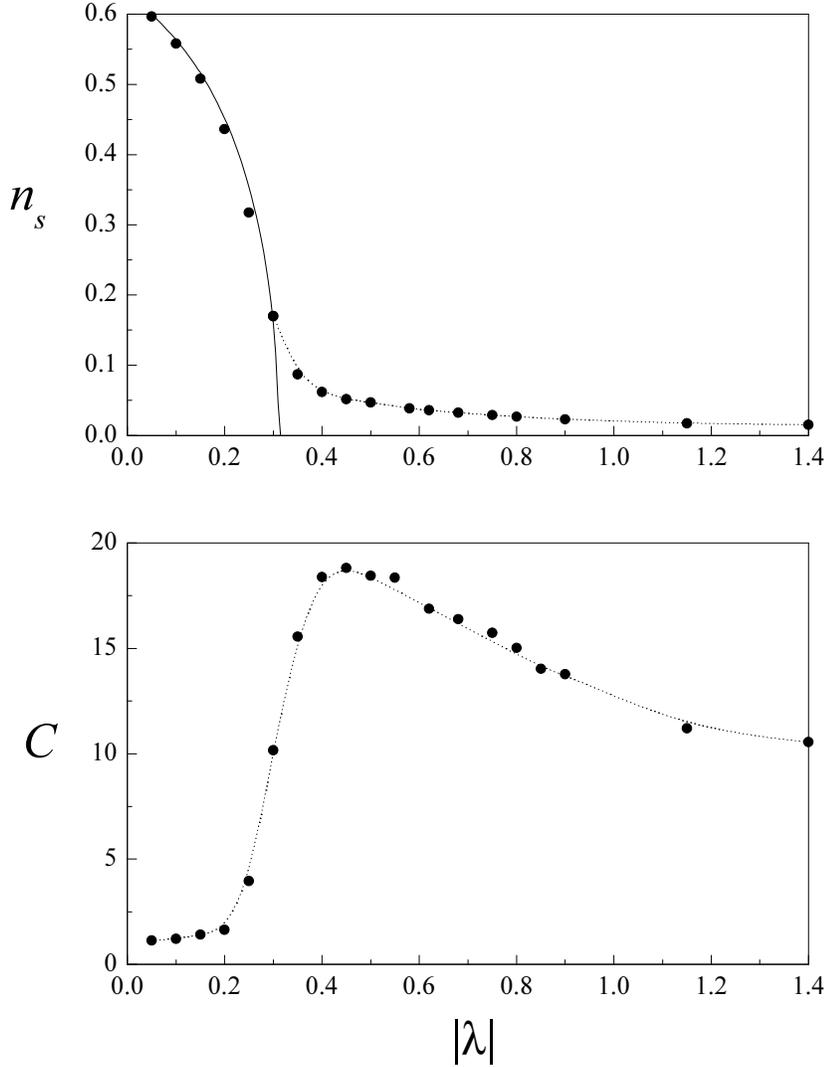}}
\caption{Upper panel: Fraction $n_s$ of synchronized oscillators as
a function of $|\lambda|$, for $K_{\max}=4$. The full curve is the
analytical prediction, and the dotted curve has been added as a
guide to the eye in the clustering regime. Lower panel: The
corresponding number of clusters, $C$. The dotted curve is a spline
approximation.} \label{fig5}
\end{figure}

\begin{figure} \centering
\resizebox{.8\columnwidth}{!}{\includegraphics{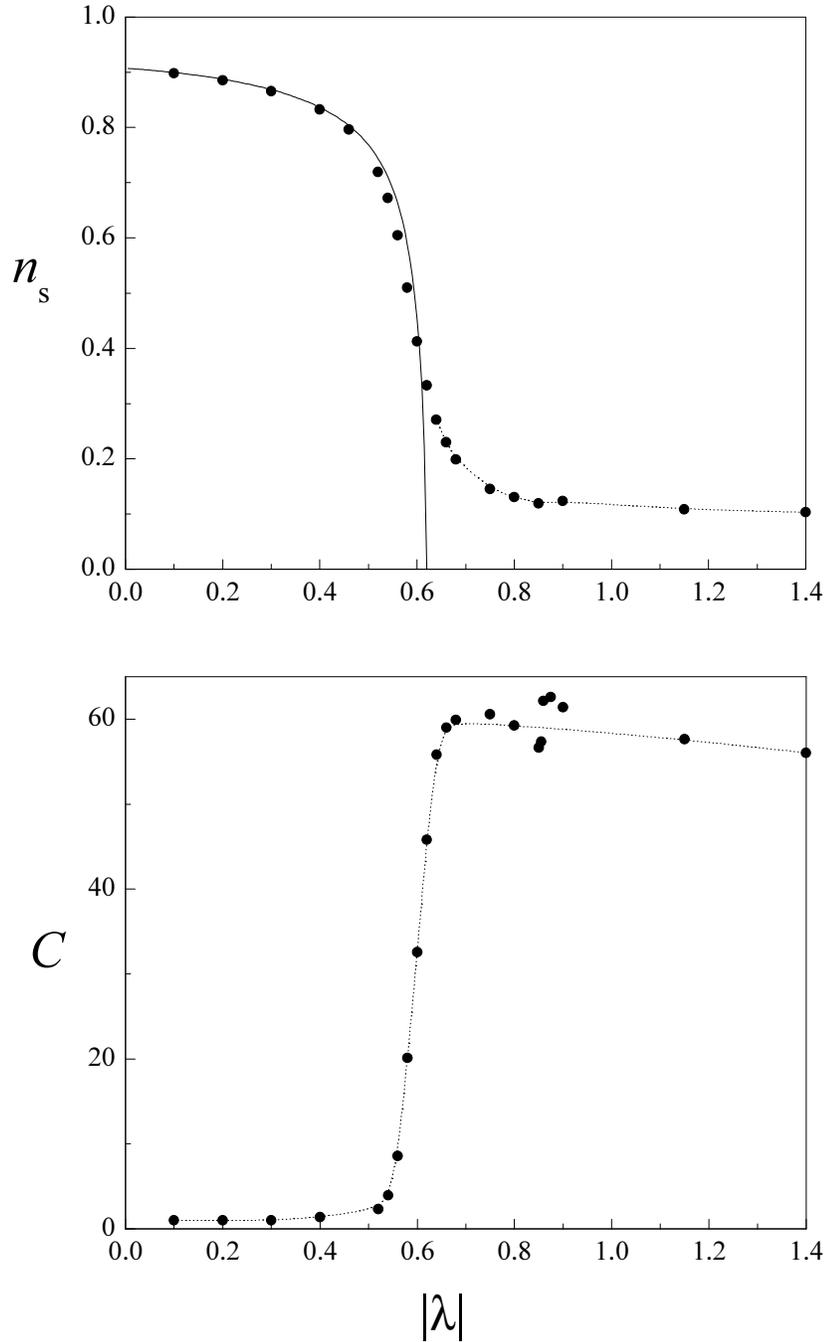}}
\caption{As in Fig. \ref{fig5}, for $K_{\max}=10$.} \label{fig6}
\end{figure}

A more quantitative illustration of this dependence on $K_{\max}$ is
provided by the numerical results presented in Figs. \ref{fig5} and
\ref{fig6}. The upper panel of Fig. \ref{fig5} shows the fraction
$n_s$ of synchronized oscillators as a function of $|\lambda|$, for
$K_{\max}=4$. The curve corresponds to the analytical prediction.
For $|\lambda|$ larger than the critical value, $n_s$ decreases
rapidly, such that only $2 \%$ of the ensemble remains clustered for
$|\lambda| \approx 1$. In the lower panel, we plot the corresponding
number of clusters, $C$. Below the transition, in the
synchronization region, $C\approx 1$. Around the transition point,
the number of clusters increases abruptly, and reaches a maximum
just below $C = 20$, before beginning to decline for larger
$|\lambda|$.

For $K_{\max}=10$, the behaviour of both $n_s$ and $C$ is
qualitatively similar, but important quantitative differences are
apparent. First, the fraction of synchronized oscillators in the
clustering region is much higher and its decay is slower, with
approximately $10 \%$ of the ensemble still entrained into clusters
for $|\lambda|\approx 1$. Second, the number of clusters at the
transition grows above $C =60$, and the subsequent decay is
extremely slow. For values of $|\lambda|$ twice as large as the
critical point, $C$ remains practically as large as just above the
transition.

\section{Conclusion}

In this paper, we have considered an ensemble of coupled phase
oscillators where the strength of the interaction is generally
different for each oscillator pair. This heterogeneity is
represented by symmetric positive interaction coefficients $W_{ij}$
and, together with the distribution of natural frequencies
$\omega_i$, adds diversity to the ensemble. Our main goal has been
to show that Kuramoto's theory for the synchronization transition of
an infinitely large ensemble of globally coupled phase oscillators
can be extended to the case of heterogeneous coupling, yielding
exact results when the interaction coefficients can be factorized as
$W_{ij}=k_i q_j$. The factors $k_i$ and $q_i$ turn out to be,
respectively, the coupling strength of oscillator $i$ with the mean
field generated by the ensemble, and the weight with which the same
oscillator contributes to that mean field. The three attributes that
characterize each oscillator, $\omega_i$, $k_i$, and $q_i$, are
distributed over the ensemble according to a function $G(\omega,k,q
)$ which, in general, introduces statistical correlations between
them.

In the analysis of our results, we have paid particular attention to
the effects of correlations between natural frequencies $\omega_i$,
couplings $k_i$, and weights $q_i$. As it may  have been expected,
if the couplings and/or the weights of the oscillators which trigger
synchronization --around the center of the frequency distribution--
are relatively small, an overall larger coupling will be necessary
to effectively have synchronized states. The effect of correlations
between couplings and weights, on the other hand, is less obvious.
We have found that a sufficiently strong anti-correlation between
$k_i$ and $q_i$ is able to completely inhibit synchronization, in
the sense that synchronized states are suppressed even in the
presence of arbitrarily large couplings. In other words, if the mean
field is strongly dominated by oscillators whose coupling is weak,
and {\it vice versa}, the ensemble may be unable to display coherent
collective oscillations.

As a validation of the analytical formulation, we have performed
numerical calculations for ensembles of $10^4$ oscillators, with
several combinations of the distribution of couplings and weights.
Numerical and analytical results are in good overall agreement,
except for a systematic departure in the parameter regions where the
synchronization transition takes place. In these zones, numerical
results reveal the existence of a regime of partial synchronization,
where oscillators become entrained into several internally
synchronized clusters. As the transition is approached, these
clusters grow in size and, at the same time, they progressively
aggregate with each other. Eventually, they collapse into a single
cluster, which can be identified with the macroscopic fraction of
synchronized oscillators predicted by the analytical formulation.
Previous numerical results on a special case of the present system
\cite{nos}, seem to indicate that frequency clustering is a
finite-size effect which disappears for infinitely large systems
--where, precisely, the analytical approach is formulated. In finite
ensembles, clusters may emerge from localized fluctuations in the
distribution of frequencies and couplings, which trigger the mutual
synchronization of a few oscillators.

The regime of clustering preceding the transition to synchronization
found in our system, is reminiscent of a similar phenomenon
well-known to occur in ensembles of coupled chaotic elements
\cite{clus1,clus2,clus3,mmz}. In the present case, heterogeneity in
the ensemble replaces chaotic dynamics as the factor which
counteracts the effects of coupling and gives origin to the
transition. In spite of the fact that it may disappear in the
thermodynamical limit, clustering remains the richest dynamical
regime of oscillator ensembles because of its complexity and
diversity \cite{conhudson}.

In connection with chaotic systems, it would be interesting to study
whether the introduction of coupling heterogeneity in ensembles of
chaotic dynamical elements have effects similar to those described
here for the synchronization transition of phase oscillators. More
complex individual dynamics are also expected to give rise to new
collective phenomena, such as amplitude effects. Distributed
couplings $k_i$ and weights $q_i$ can be straightforwardly
incorporated to standard models of interacting chaotic systems
\cite{mmz}. For instance, linear heterogeneous coupling between $N$
chaotic elements whose individual dynamics is governed by the
equation $\dot {\bf x} ={\bf f} ({\bf x})$ may be introduced as
\begin{equation}
\dot {\bf x}_i ={\bf f} ({\bf x}_i) + k_i ({\bf X}-{\bf x}_i),
\end{equation}
$i=1, \dots, N$, with ${\bf X} =N^{-1} \sum_i q_i {\bf x}_i $.
Finally, let us point out that Kuramoto's theory has been extended,
at least partially, to study the synchronization properties of
ensembles formed by other kinds of dynamical elements, in
particular, by active rotators \cite{ar1,ar2}. Consideration of such
ensembles with the kind of heterogeneous coupling studied here would
be a natural continuation of the present work.

\end{document}